\documentclass[a4paper]{article}
\usepackage{amsmath,amsfonts,amssymb,amsthm} 

\newtheorem{theor}{Theorem}
\theoremstyle{definition}
\newtheorem*{lemma}{Lemma}
\newtheorem{state}[theor]{Proposition}
\newtheorem*{cor}{Corollary}
\theoremstyle{remark}
\newtheorem{rem}{Remark}

\newcommand{\BBR}{\mathbb{R}}
\newcommand{\BBC}{\mathbb{C}}
\newcommand{\BBN}{\mathbb{N}}

\newcommand{\dd}{\partial}
\newcommand{\vph}{\varphi}
\newcommand{\Id}{{\mathrm d}}
\newcommand{\cD}{{\mathcal{D}}}

\newcommand{\teta}{{\tilde{\eta}}}

\newcommand{\by}[1]{\textit{{#1}}}
\newcommand{\jour}[1]{\textit{{#1}}}
\newcommand{\vol}[1]{\textbf{{#1}}}
\newcommand{\book}[1]{\textrm{{#1}}}

\title
{Hirota's virtual multi\/-\/soliton solutions of
$N=2$~supersymmetric KdV 
equations}

\author{Arthemy V. Kiselev%
\thanks{
Mathematical Institute, University of Utrecht,
P.O.Box 80.010, 3508\,TA Utrecht, The Netherlands;
 \textit{E-mail}: \texttt{A.V.Kiselev\symbol{"40}uu.nl}}
\thanks{\textit{Present address}:
Institut des Hautes \'Etudes Scientifiques,
Le Bois-Marie, 35~Route de Chartres,
F-91440 Bures-sur-Yvette, France.}{\ },
V\'eronique Hussin%
\thanks{
D\'epartement de Math\'ematiques et de Statistique,
Universit\'e de Montr\'eal,
C.P.~6128, succ.\ Centre\/-\/ville, Montr\'eal, Qu\'ebec H3C~3J7, Canada;  \textit{E-mail}: 
\texttt{hussin\symbol{"40}dms.umontreal.ca}}
}

\date{October 6, 2008; in final form December 1, 2008}

\begin{document}
\maketitle

\begin{abstract}
We prove that Mathieu's $N=2$ supersymmetric Korteweg\/--\/de Vries equations
with $a=1$ or~$4$ admit Hirota's $n$\/-\/supersoliton solutions,
whose nonlinear interaction does not produce any phase shifts.
For initial profiles that can not be distinguished from a one\/-\/soliton
solution 
at times $t\ll0$, we reveal the possibility of a spontaneous decay and,
within a finite time, transformation into a solitonic solution with a different
wave number. This paradoxal effect is realized by the completely integrable
$N=2$ super\/-\/KdV systems, whenever the initial soliton is loaded with
other solitons that are virtual and become manifest through the 
$\tau$\/-\/function as the time grows.

\end{abstract}

\paragraph*{Introduction.}
Let $u(x,t;\theta_1,\theta_2)$ be a scalar 
complex bosonic $N=2$ superfield.
We consider Mathieu's 
$N=2$ supersymmetric Korteweg\/--\/de Vries
equations~\cite{MathieuTwo,MathieuNew}
\begin{equation}\label{SKdV}
u_t=-u_{xxx}+3\bigl(u\cD_1\cD_2u\bigr)_x+\tfrac{a-1}{2}\bigl(\cD_1\cD_2u^2\bigr)_x
 +3au^2u_x,\qquad
 \cD_i=\frac{\dd}{\dd\theta_i}+\theta_i\cdot\frac{\dd}{\dd x}
\end{equation}
with $a=4$ or $a=1$. In this paper, we construct $n$-\/supersoliton solutions
of~\eqref{SKdV},
\begin{equation}\label{HirotaForm}
u=A\cdot \cD_1\cD_2\log\tau_{k_1,\ldots,k_n}(x,t;\theta_1,\theta_2),\qquad
A(a)=\text{const}\in\BBR,
\end{equation}
that 
possess the following properties. 
First, the linear truncations of Hirota's exponential series,
\begin{equation}\label{ZeroInteraction}
\tau=1+\sum_{i=1}^n \alpha_i\exp\eta_i,\qquad
  \eta_i=k_ix+\omega(k_i)\cdot t+\lambda(k_i)\cdot\theta_1\theta_2,
  \quad k_i\in\BBR,\ \alpha_i\in\BBR,
\end{equation}
yield \emph{exact} solutions of~\eqref{SKdV}.
Functions~\eqref{ZeroInteraction} contain no pairwise
interaction
($A_{ij}(k_i,k_j)\equiv0$, see~\eqref{TauSoliton} below) 
and no multiple\/-\/order interaction terms 
(respectively, $A_{i_1 \cdots i_\ell}\bigl(k_{i_1}$, $\ldots$, $k_{i_\ell}\bigr)
 =\prod_{i_\alpha<i_\beta} 
   A_{i_\alpha i_\beta}\bigl(k_{i_\alpha},k_{i_\beta}\bigr)\equiv0$,
see~\cite{JimboMiwaDate}).
Second, as $n$~different solitons constitute an $n$-\/soliton solution and
obey the nonlinear superposition formulas while overlapping, they
acquire no phase shifts having become distant from each other. 
Thus the spatially extensive solitary waves demonstrate a dimensionless
behaviour of particles with elastic nonlinear collisions.
We recall that a similar
example of the field representation for point particles
was known, e.g., for soliton solutions of the integro\/-\/differential
Benjamin\/--\/Ono equation~\cite{Blaszak}.

Third, we show that 
a fast soliton
that precedes a set of other waves always decays spontaneously and,
sooner or later, the observed solution
converts into the slowest soliton from that set. We say that 
the initial soliton, which was visible at $t\ll0$, was loaded with $n-1$
virtual solitons ahead of it. 
We demonstrate how   
the fastest soliton gains
the virtual ones and becomes virtual itself, while the slowest virtual soliton
manifests itself behind the others and becomes visible.
Of course, these properties of the new solutions for $N=2$ super\/-\/KdV
equations~\eqref{SKdV} are radically different from superposition laws and
phase shifts that are exhibited by the $n$-\/solitons for the standard KdV
equation, see~\eqref{KdV} below, and by the super\/-\/soliton 
solutions~\eqref{KdVtoSKdV} derived from them.
This effect, which is new to the best of our knowledge, may be useful 
in verifying the relevance of nonlinear $N=2$ supersymmetric KdV 
equations~\eqref{SKdV} in field theory and, more generally, establishing 
the physical sense of the Grassmann variables $\theta_1$,~$\theta_2$.

The paper is organized as follows. First, we summarize well\/-\/known
properties of $n$-\/soliton solutions for the 
KdV equation. Here we also recall how these
solutions are inherited by Hamiltonian $N=2$ super\/-\/KdV 
systems~\eqref{SKdV} from the purely bosonic~KdV. 
Then we formulate our main result, describing the 
$\tau$-\/func\-ti\-ons and dispersion relations in the new $n$\/-\/supersolitons
for~\eqref{SKdV}. If $a=1$, then the new super\/-\/fields yield exact
$n$\/-\/soliton solutions of the Krasil'shchik\/--\/Ker\-s\-ten 
equation~\cite{JKKerstenEq}.
We establish the paradoxal properties of the $n$-solitons for~\eqref{SKdV},
including their spontaneous decay followed by transition into 
 virtual states. 

\paragraph*{\S1. Known $n$-\/solitons for the $N=2$ super\/-\/KdV.}
Some classes of exact travelling wave solutions
$u(x+ct+i\theta_1\theta_2$, $\theta_1+i\theta_2)$ for
equations~\eqref{SKdV} with $a=-2$,\ $\pm1$,\ $\pm4$ were obtained
in~\cite{AyariVHPW} using the symmetry invariance technique. First, a
Lie subgroup of point symmetries for~\eqref{SKdV} with $a\in\BBR$ was
chosen such that the respective group invariants specified the
arguments of the solutions~$u$ as above.
It follows easily that the Taylor expansion 
$u=\rho_0+(\theta_1+i\theta_2)\cdot \rho_1$ in the second argument,
with both $\rho_0$ and $\rho_1$ depending on $x+ct+i\theta_1\theta_2$,
converts~\eqref{SKdV} to 
the triangular system of ODE upon $\rho_0$ and~$\rho_1$.
The second assumption of~\cite{AyariVHPW}
is that solutions of the first equation in this system, which is
$\rho_0''-a\rho_0^3+i(a+2)\rho_0\rho_0'+c\rho_0+\text{const}=0$,
have no movable critical points.    
This yields the admissible values
$a=-2,\pm1,\pm4$. Particular solutions
$\rho_0(x+ct+i\theta_1\theta_2)$, $\rho_1\equiv0$, which are
found in~\cite[Eq.~(3.15)]{AyariVHPW}
for~\eqref{SKdV} with $a\in\{1,4\}$, are close to the
one\/-\/solitons~\eqref{HirotaForm} with $c=\omega/k$ 
in~\eqref{ZeroInteraction}.
However, the pairwise or collective interaction of the travelling
waves was not discussed in~\cite{AyariVHPW}.

At the same time, it is clear that the $N=2$ super\/-\/KdV
equations~\eqref{SKdV}, and not only with the values $a\in\{-2,1,4\}$
that ensure the complete integrability, 
possess multi\/-\/soliton solutions. 
Indeed, expand the bosonic $N=2$ super\/-\/field $u$ in $\theta_1$ 
and~$\theta_2$,
\begin{equation}\label{N2SuperField}
u=u_0+\theta_1 u_1+\theta_2 u_2+\theta_1\theta_2 u_{12},
\end{equation}
and substitute 
\eqref{N2SuperField} in~\eqref{SKdV}. 
Then, in particular, the even components $u_0$ and~$u_{12}$ satisfy the
system~\cite{MathieuTwo,MathieuNew,JKKerstenEq}
\begin{subequations}\label{JKKerstenBL}
\begin{align}
u_{0;t}&+u_{0;xxx}-3au_0^2u_{0;x}+(a+2)(u_0u_{12})_x-(a-1)(u_1u_2)_x=0,
\label{GetmKdV}\\
u_{12;t}&+u_{12;xxx}+6u_{12}u_{12;x}-(a+2)u_0u_{0;xxx}
-3a\bigl(u_{0;x}u_{0;xx}+u_0^2u_{12;x}+2u_0u_{0;x}u_{12}\bigr)\notag\\
{}&{}-3u_2u_{2;xx}-(a+2)u_1u_{1;xx}+6a\bigl(u_0u_1u_{2;x}
  -u_0u_{1;x}u_2+u_{0;x}u_1u_2\bigr)=0.\label{GetKdV}
\end{align}
\end{subequations}
Set the odd components $u_1=u_2\equiv0$. First, if $a=-2$, then 
system~\eqref{JKKerstenBL} splits 
to the modified KdV equation~\eqref{GetmKdV} upon~$u_0$ together with 
the KdV\/-\/type equation~\eqref{GetKdV} with a feed\/-\/back. In what follows,
we do not consider the well\/-\/studied case~$a=-2$.
If, under the same assumption $u_1=u_2=0$, we let~$a=1$, 
then system~\eqref{JKKerstenBL} is the Krasil'shchik\/--\/Kersten
coupled KdV\/-\/mKdV equation~\cite{JKKerstenEq}, see 
also~\cite{Ismet,HonFan,Hong}. In the next section we
describe a new class of its multi\/-\/soliton solutions in Hirota's form.

Conversely, set the components  $u_0$,\ $u_1$,\ and~$u_2$ to zero.
Then for any $a\in\BBR$ the entire $N=2$ supersymmetric equation~\eqref{SKdV}
reduces to the purely bosonic, scalar KdV equation
\begin{equation}\label{KdV}
u_{12;t}+u_{12;xxx}+6u_{12}u_{12;x}=0.
\end{equation}
Therefore the super\/-\/function
\begin{equation}\label{KdVtoSKdV}
u=\theta_1\theta_2\cdot u_{12}(x,t),
\end{equation}
where $u_{12}$ satisfies~\eqref{KdV}, is a solution of~\eqref{SKdV}.
Reciprocally, each function~\eqref{KdVtoSKdV} yields the solution
$u_{12}=\cD_2\cD_1(u)$ of~\eqref{KdV}. Note that for the class~\eqref{KdVtoSKdV}
one transfers the formulas of B\"acklund autotransformation for KdV 
to the $N=2$ super\/-\/KdV equations~\eqref{SKdV}.

Substitution~\eqref{KdVtoSKdV} further implies that equations~\eqref{SKdV}
inherit the classical Hirota $n$-\/solitons~\eqref{HirotaForm}
from the solutions
\begin{equation}\label{KdVSoliton}
u_{12}=A(a)\cdot{\Id^2}/{\Id x^2}\,\log\tau
\end{equation}
of the KdV equation~\eqref{KdV}, see~\cite{JimboMiwaDate}, where
the $\tau$-\/function is
\begin{equation}\label{TauSoliton}
\tau=\sum_{\substack{\mu_\ell=0,1\\ 1\leq\ell\leq n}}\exp\Bigl(
 \sum_{\ell=1}^n \mu_\ell\cdot(\eta_\ell+\log\alpha_\ell)
 +\sum_{1\leq\ell<m\leq n}\mu_\ell\mu_m\cdot\log A_{\ell m}
\Bigr),
\end{equation}
and the dispersion and interaction (now $\lambda\equiv0$) are given through
\begin{equation}\label{FromKdV}
A(a)\equiv2,\qquad \omega=-k^3,\qquad 
A_{\ell m}={(k_\ell-k_m)^2}\bigr/{(k_\ell+k_m)^2},\qquad \alpha_\ell\in\BBR.
\end{equation}

\begin{lemma}
After each pairwise interaction, which is encoded by $A_{\ell m}(k_\ell,k_m)$
in~\eqref{FromKdV}, the solitons for KdV equation~\eqref{KdV} acquire the phase
shifts, whose absolute values depend on the wave numbers $k_\ell$,\ $k_m$ and equal
\begin{equation}\label{PhaseShift}
\Delta\vph_{\ell m}=2\log\Bigl({|k_\ell-k_m|}\bigr/{|k_\ell+k_m|}\Bigr);
\end{equation}
the signs of the shifts
themselves are opposite for the $\ell$-th and $m$-th
solitons.
For instance, if $k_\ell>k_m>0$, then the fast soliton with the wave 
number~$k_\ell$ is pushed forward, 
and the slow soliton given by~$k_m$ retards back. 
As $t\to-\infty$, 
the two solitons move freely, still staying (respectively,
as $t\to+\infty$, having already
become) distant from each other.
\end{lemma}

The knowledge of this effect dates back to the beginning of the KdV-\/boom in late 60's, 
see~\cite{JimboMiwaDate} and references therein. 
We are obliged to re\/-\/derive formula~\eqref{PhaseShift} in order
to compare it with the application of
the same reasoning to the new solutions~\eqref{nSoliton}, see below,
of $N=2$ super\/-\/equations~\eqref{SKdV}. The new super\/-\/solitons
demonstrate a principally
different behaviour both from the analytic and physical points of view.

\begin{proof}
Formula~\eqref{PhaseShift} is particularly transparent on the level of the 
$\tau$-\/functions~\eqref{TauSoliton}; without loss of generality, 
we assume $n=2$ and thus let
$\ell=1$, $m=2$ for a two\/-\/soliton solution of~\eqref{KdV} given by
$k_1>k_2>0$. 
Let $x_1(t)$ and $x_2(t)$ be the coordinates of the peaks of the first and second solitons, respectively, at the times $t$ when they are sufficiently distant from each other.
For convenience, we set all $\alpha_i\mathrel{{:}{=}}1$, 
and hence in view of~\eqref{TauSoliton} 
the $\tau$-\/function has the form
\[
\smash{\tau_{k_1k_2}(x,t)}=1+\smash{\exp\eta_1+\exp\eta_2
   +A_{12}(k_1,k_2)\cdot\exp(\eta_1+\eta_2)}.
\]
First, let $t\ll0$ be fixed. Consider the fast first soliton, which has not yet
overpassed the slow second one. In the vicinity of its peak located 
at~$x_1(t)$, we have $\eta_1\approx0$ and, since the peak of the second
soliton is still far ahead at $x_2(t)\gg x_1(t)$ for $t\ll0$, we deduce
$\eta_2(x)\ll0$ for $x\approx x_1(t)$. Hence,\footnote{Here and everywhere
below the standard symbol $\rightrightarrows$ denotes the uniform convergence
of a functional sequence on a finite closed interval. 
The symbol~$\downdownarrows$ denotes the uniform convergence from above
(in our case, to the zero function).}
\begin{equation}\label{KdV4Cases1st}
\smash{\tau_{k_1k_2}(x,t)}=1+\smash{\exp\eta_1+\exp\eta_2\cdot\bigl(1+A_{12}\cdot\exp\eta_1\bigr)
   \rightrightarrows 1+\exp\eta_1=\tau_{k_1}(\eta_1)},
\end{equation}
because $\exp\eta_2(x)\downdownarrows0$ uniformly as $t\to-\infty$
on any finite interval that contains~$x_1(t)$. The asymptote is just the fast
soliton in its pure form~\eqref{TauSoliton} without any phase shift yet.

Next, consider a vicinity $\{x\approx x_2(t)\}$ of the peak of the slow
soliton, which still goes far ahead the fast one. Now $\eta_2\approx0$ and,
since $x_2(t)\gg x_1(t)$ for $t\ll0$, we have $\eta_1(x)\gg0$ at 
any~$x\approx x_2(t)$. Consequently,
\[
\tau_{k_1k_2}(x,t)=\exp\eta_1\cdot\smash{\Bigl(1+\exp\bigl(\eta_2+\log A_{12}\bigr)
   +\exp(-\eta_1)\cdot\bigl(1+\exp\eta_2\bigr)\Bigr)}.\mathstrut
\]
Acting by the operator $2\Id^2/\Id x^2\circ\log$ and noting that the 
phase~$\eta_1$ is only linear in~$x$
(it does not contain higher powers of $x$, we conclude that
\begin{multline*}
u_{k_1k_2}(x,t)=2\frac{\Id^2}{\Id x^2}\Bigl[
 \bigl(k_1 x+\omega_1 t\bigr) +
 \log\Bigl(1+\exp\bigl(\eta_2+2\log\tfrac{k_1-k_2}{k_1+k_2}\bigr)
            +\exp(-\eta_1)\cdot\bigl(1+\exp\eta_2\bigr)
     \Bigr)
\Bigr]\\
{}\rightrightarrows u_{k_2}\bigl(\eta_2+2\log\tfrac{k_1-k_2}{k_1+k_2}\bigr),
\end{multline*}
since $\exp(-\eta_1)\downdownarrows0$ on a finite interval around~$x_2(t)$ 
as~$t\to-\infty$. The negative phase shift is precisely~\eqref{PhaseShift};
as $t\to+\infty$, it will vanish, and hence the overall phase shift of the slow
soliton will be positive.

Notice that, initially,
the slow soliton is displaced forward by~$|\Delta\vph_{k_1k_2}|$
with respect to an identical soliton~\eqref{KdVSoliton} with the same wave
number~$k_2$
that would arrive to the origin $x_2(t)=0$ at~$t=0$. Thus, we recall, the
system of coordinates $(x,t)$ used in Hirota's method is such that the center
of mass for the two soliton solution achieves $x=0$ at some~$t<0$.

\smallskip
Second, suppose $t\to+\infty$; by now, the fast soliton has left the slow one
far behind. Therefore, in a vicinity $\{x\approx x_2(t)\}$ of the slow peak 
with $\eta_2(x)\approx0$, we have $\eta_1(x)\ll0$ at a fixed time~$t$, whence
\[
\smash{\tau_{k_1k_2}(x,t)}=1+\smash{\exp\eta_2+\exp\eta_1\cdot\bigl(1+A_{12}\exp\eta_2\bigr)
   \rightrightarrows 1+\exp\eta_2=\tau_{k_2}(\eta_2)},
\]
because $\exp\eta_1(x)\downdownarrows0$ 
near~$x_2(t)$ as~$t\to+\infty$.

Concentrated around $x_1(t)$ such that $\eta_1=0$, the fast soliton 
shifts forward, borrowing the energy and momentum from the slowly moving
obstruction. Indeed, by $\eta_2\bigl(x_1(t)\bigr)\gg0$ we have
\[
\tau_{k_1k_2}(x,t)=\exp\eta_2\cdot
  \smash{\Bigl(1+\exp\bigl(\eta_1+\log A_{12}\bigr)+\exp(-\eta_2)\cdot
  \bigl(1+\exp\eta_1\bigr)\Bigr)},\mathstrut
\]
which implies
\begin{multline}\label{KdV4CasesLast}
u_{k_1k_2}(x,t)=2\frac{\Id^2}{\Id x^2}\Bigl[\bigl(k_2 x+\omega_2 t\bigr)
 +\log\Bigl(1+\exp\bigl(\eta_1+2\log\tfrac{k_1-k_2}{k_1+k_2}\bigr)
      +\exp(-\eta_2)\cdot\bigl(1+\exp\eta_1\bigr)\Bigr)\Bigr]\\
{}\rightrightarrows u_{k_1}\bigl(\eta_1+2\log\tfrac{k_1-k_2}{k_1+k_2}\bigr)
\end{multline}
on $\{x\approx x_1(t)\}$ as~$t\to+\infty$, because 
$\exp(-\eta_2)\downdownarrows0$ there. Comparing the in\/-{}
and out\/-\/going states of the two solitons, we recognize the accumulated
phase shifts $\pm\Delta\vph_{k_1k_2}$ for
the first and second solitons, respectively.
\end{proof}

\begin{rem}\label{RemFermiConst}
The dispersion law in~\eqref{FromKdV} is determined by the Hirota
bilinear equation for the KdV equation~\eqref{KdV}, see~\cite{JimboMiwaDate},
\[
D_x(D_t+D_x^3)\,\tau\cdot\tau=0,
\]
where $D_x$ and $D_t$ are Hirota's derivatives.
We recall that 
a bilinear representation for the Manin\/--\/Radul $N=2$ supersymmetric
KdV equation~\eqref{SKdV} with $a=-2$, see~\cite{ManinRadul}, was
obtained in~\cite{GhoshSarmaIK} for its expression in
Inami\/--\/Kanno's form~\cite{InamiKanno}. 
The bilinear representation for the $a=-2$ 
super\/-\/KdV~\eqref{SKdV} is such that the $\tau$-\/functions
for the $n$\/-\/soliton solutions, assigned to the wave numbers $k_1$,
$\ldots$, $k_n$, contain the Grassmann constants $\zeta_1$, $\ldots$,
$\zeta_n$. 
\end{rem}

\paragraph*{\S2. New (virtual) $n$-\/solitons for the $N=2$ super\/-\/KdV.}
We discard the idea of using any `Grassmann constants,'   
see Remark~\ref{RemFermiConst} above. Consequently,
the most general admissible form of the Hirota exponentials is
\begin{equation}\label{HirotaExp}
\eta=kx+\omega t+\lambda\theta_1\theta_2,
\end{equation}
where $k\in\BBR$ are the wave numbers, $\omega$~are the frequencies,
and $\lambda\in\BBC$ describes the dependence on the
Grassmann variables $\theta_1$,~$\theta_2$ for 
$n$-\/supersolitons $u_{k_1\cdots k_n}(x,t;\theta_1,\theta_2)$. 

\begin{state}\label{nSolState}
{\renewcommand{\theenumi}{\roman{enumi}}
\begin{enumerate}
\item\label{FirstOneSoliton}
The travelling wave solutions with Hirota's
exponentials~\eqref{HirotaExp} for the $N=2$ supersymmetric KdV
equation~\eqref{SKdV} with $a=1$ or $a=4$ are
\begin{equation}\label{OneSoliton}
u=A\cdot \cD_1\cD_2\log\bigl(1+\alpha\exp\eta\bigr),
\end{equation}
where the dispersion and the relation between the wave
number~$k$ and the \emph{nonzero} factor~$\lambda$ are
\begin{equation}\label{Specify}
\omega=-k^3,\quad \lambda=\pm ik,\quad k\in\BBR, \quad \alpha\in\BBR;
\qquad \left\{\begin{aligned}A&=1\text{ if }a=1,\\
A&=\tfrac{1}{2}\text{ if }a=4.\end{aligned}\right.
\end{equation}
\item
The Hirota two\/-\/soliton solutions of~\eqref{SKdV} with $a\in\{1,4\}$
and $\lambda\neq0$ in~\eqref{HirotaExp},
\[
u=A\cdot \smash{\cD_1\cD_2\log\bigl(1+\alpha_1\exp\eta_1 +\alpha_2\exp\eta_2
 +A_{12}\alpha_1\alpha_2\exp(\eta_1+\eta_2)\bigr)},
\]
exist if and only if relations~\eqref{Specify} for Hirota's
exponentials~\eqref{HirotaExp} hold and there is no
coupling between the exponentials for one\/-\/soliton 
solutions~\eqref{OneSoliton}:\ ${A_{12}(k_1,k_2)\equiv0}$.
\item\label{ThirdnSoliton}
The exact $n$-supersoliton solution of the $N=2$ super\/-\/KdV \eqref{SKdV}
with $a\in\{1,4\}$ and $\lambda_i\neq0$ is
\begin{equation}\label{nSoliton}
u=-A\cdot\frac{\sum_{i=1}^n\lambda_i\alpha_i\exp\tilde{\eta}_i}{1
  +\sum_{i=1}^n\alpha_i\exp\tilde{\eta}_i}
 +A\cdot\theta_1\theta_2\frac{\Id}{\Id x}\left[
   \frac{\sum_{i=1}^n k_i\alpha_i\exp\tilde{\eta}_i}{1
  +\sum_{i=1}^n\alpha_i\exp\tilde{\eta}_i}\right].
\end{equation}
Here $n\in\BBN$ is arbitrary and we put the reduced phase be
\begin{equation}\label{ReducedPhase}
\tilde\eta=kx+\omega t;
\end{equation}
then, substituting~\eqref{Specify}
for $A(a)$, $\omega(k)$, and $\lambda(k)$, we expand the
supersoliton~\eqref{HirotaForm} with
the $\tau$-\/function 
\[
\tau_{k_1\cdots k_n}(x,t;\theta_1,\theta_2)=
    1+\smash{\sum\nolimits_{i=1}^n \alpha_i\exp\teta_i\cdot(1+\lambda_i\theta_1\theta_2)}
\]
in $\theta_1\theta_2$, whose square is zero.
\item\label{ExhaustSt}
The $a=-2$ super\/-\/equation~\eqref{SKdV}
does not possess travelling wave
solutions~\eqref{OneSoliton} other than~\eqref{FromKdV} 
with $\lambda\equiv0$, and hence it does
not admit 
$n$-\/supersolitons~\eqref{nSoliton}
for any~$n\geq1$.
This also holds for any other $a\in\BBR\setminus\{1,4\}$,
which still do admit the $n$-\/solitons given 
by~(\ref{KdVtoSKdV}--\ref{FromKdV}) but not 
by~(\ref{OneSoliton}--\ref{Specify}).
\end{enumerate}%
}
\end{state}

\begin{proof}
Once (\ref{OneSoliton}--\ref{nSoliton}) are known, 
statements~\ref{FirstOneSoliton}--\ref{ThirdnSoliton} are verified
by direct calculation. The exhaustive conclusion of
statement~\ref{ExhaustSt} is obtained using
analytic environment~\cite{SsTools}.      
\end{proof}

\begin{rem}
In~\cite{Ibort}, a particular class of exact solutions was constructed for
the Manin\/--\/Radul $N=2$ super\/-\/KP equation and for its reduction to the
$N=2$ super\/-\/KdV equation~\eqref{SKdV} with~$a=-2$. The components of these
solutions (those are called \emph{solitinos} in~\textit{loc.\ cit.}) are 
at most quadratic in Hirota's exponentials $\exp\eta_i$ for any number~$n\geq2$
of interacting solitinos, see~\cite[Eqs~(39--40)]{Ibort}. 

We have extended this result onto the entire triple $a\in\{-2,1,4\}$ of
integrable Hamiltonian super\/-\/equations~\eqref{SKdV}. Namely, 
for $N=2$ super\/-\/KdV equations with $a=1$ or~$4$ we revealed 
the $n$\/-\/soliton solutions~\eqref{nSoliton}, whose $\tau$-\/functions
remain linear in $\exp\eta_i$ for any~$n\geq1$.
We stress that the unexpected $n$-\/solitons~\eqref{nSoliton} of
the $N=2$ super\/-\/KdV must also solve the $N=2$ supersymmetric 
KP~equations that reduce to~\eqref{SKdV} with $a\in\{1,4\}$. Definitely,
the admissible super\/-\/systems are not the Manin\/--\/Radul $N=2$
super\/-\/KP~\cite{ManinRadul} that implies~$a=-2$.

Finally, we note that the constraint $u_1=u_2\equiv0$ is fulfilled for the
components of super\/-\/fields~\eqref{nSoliton}. Therefore 
Proposition~\ref{nSolState} yields exact solutions of the
Krasil'shchik\/--\/Kersten equation, see~\eqref{JKKerstenBL}, whenever~$a=1$.
Likewise, it provides the $n$\/-\/solitons for the bosonic limit 
of~\eqref{SKdV} if~$a=4$.
\end{rem}

Under the nonlinear superposition of $n$~solitons, exact 
solutions~\eqref{nSoliton} of super\/-\/equation \eqref{SKdV},
which are given by linear truncations~\eqref{ZeroInteraction} of
Hirota's $\tau$-\/functions, demonstrate the properties that are essentially
different from the standard picture~\eqref{PhaseShift} for KdV~\eqref{KdV},
regarding both the asymptotes at the time infinity and the phase shifts for
each of the $n$~solitons. We claim that 
\begin{quote}
the asymptotic behaviour ($t\to\pm\infty$) of an $n$-\/soliton 
solution~\eqref{nSoliton}
for~\eqref{SKdV} is completely determined by the soliton that moves behind 
the others, and that no phase shifts at all are accumulated as the time~$t$
runs from~$-\infty$ to~$+\infty$.
\end{quote}

\noindent\textit{The proof} goes in parallel with 
(\ref{KdV4Cases1st}--\ref{KdV4CasesLast}), see p.~\pageref{KdV4Cases1st},
and the reasonings are alike for all $n\geq2$, hence we set $n=2$ and suppose
$k_{\text{max}}=k_1>k_2=k_{\text{min}}>0$. Without loss of generality, we
assume $\alpha_i=1$ for all~$i$. Recall that $\tau_{k_1k_2}(x,t;\theta_1,
\theta_2)=1+\exp\eta_1+\exp\eta_2$, where the full phases $\eta_i$ 
are~\eqref{HirotaExp} and the dispersion is specified by~\eqref{Specify}.
This yields the solutions~\eqref{nSoliton} of~\eqref{SKdV},
where $A(a)$ is given in~\eqref{Specify} for either~$a=1$ or~$a=4$.

Notice that the essential non\/-\/constant part of a soliton~\eqref{OneSoliton}
is, as usual, concentrated near the zero of the reduced 
phase~\eqref{ReducedPhase}. However, the first, purely imaginary summand in
the expansion~\eqref{nSoliton} does not have a peak located at 
$\tilde{\eta}\approx0$. 
At the same time, the real coefficient of $\theta_1\theta_2$ in the 
one\/-\/soliton solution does have the peak at $\teta\approx0$. 
(Let us remark that this coefficient is nothing else but one half (for $a=1$)
or a quarter (respectively, $a=4$) of the one\/-\/soliton solution~\eqref{KdVSoliton}
for~KdV, see~\eqref{FromKdV}.) 
We shall use this in what follows to refer on points~$x\in\mathbb{R}$.

So, let~$t\to-\infty$. In a vicinity of the peak $x_1(t)$ 
of the fast soliton with the
maximal wave number $k_1=k_{\text{max}}$, the reduced 
phase~\eqref{ReducedPhase} is $\teta_1\approx0$, and $\teta_2\ll0$.
For still distant solitons, $\exp\teta_2\downdownarrows0$
as $t\to-\infty$ on any finite interval $\{x\approx x_1(t)\}$. Therefore,
\begin{equation}\label{SKdV4Case1st}
\tau_{k_1k_2}(x,t;\theta_1,\theta_2)=1+\exp\eta_1+\exp\eta_2
 \rightrightarrows 1+\exp\eta_1=\tau_{k_1}(\eta_1).
\end{equation}
This is the fast soliton with its unshifted full phase~$\eta_1(x,t)$.

On the other hand, near the peak $x_2(t)$ of the slow soliton 
we have $\teta_2\approx0$ and $\teta_1\gg0$, whence
\[
\tau_{k_1k_2}(x,t;\theta_1,\theta_2)=\exp\eta_1\cdot
 \smash{\bigl(1+\exp(-\eta_1)\cdot(1+\exp\eta_2)\bigr)}.
\]
Applying the operator $A(a)\cdot \cD_1\cD_2\circ\log$ 
to the $\tau$-\/function, we obtain
the asymptote of the two\/-\/soliton solution on a finite interval 
near~$x_2(t)$:
\begin{multline*}
u_{k_1k_2}(x,t;\theta_1,\theta_2)
  =A\,\cD_1\cD_2\Bigl[\bigl(k_1 x+\omega_1 t\bigr)
 +\log\bigl(1+\lambda_1\theta_1\theta_2\bigr)
 +\log\Bigl(1+\exp(-\eta_1)\cdot\bigl(1+\exp\eta_2\bigr)\Bigr)\Bigr]\\
{}\rightrightarrows -A\lambda_1=\text{const}.
\end{multline*}
In other words, the solution near $x_2(t)$ does not vary under a small finite
motion left or right along~$x$ at~$t\ll0$. For this reason, we say that the
slow soliton has not yet manifested its presence, remaining \emph{virtual}.
We also say that the visible fast soliton, which starts behing the slower 
soliton(s) in the beginning of the elastic scattering process, 
is \emph{loaded} with the virtual soliton(s).
Note that the asymptotic behaviour 
of $u_{k_1\cdots k_n}(x,t;\theta_1,\theta_2)$ at early times $t\ll0$ is analogous for any~$n\geq2$.

\smallskip
Now consider $t\to+\infty$. The slow soliton with the wave number 
$k_2=k_{\text{min}}$ becomes visible. Near its peak at~$x_2(t)$, we have
$\teta_2\approx0$ and $\teta_1\ll0$ as soon as the fast soliton has gone
sufficiently far ahead. Consequently,
\[
\tau_{k_1k_2}(x,t;\theta_1,\theta_2)=1+\exp\eta_1+\exp\eta_2
 \rightrightarrows 1+\exp\eta_2=\tau_{k_2}(\eta_2),
\]
because $\exp\teta_1\downdownarrows0$ near~$x_2(t)$ as~$t\to+\infty$.

And the fast soliton becomes virtual. Indeed, near its peak at $x_1(t)$ such 
that $\teta_1\approx0$, we have $\teta_2\gg0$ 
and $\exp(-\teta_2)\downdownarrows0$ as $t\to+\infty$, whence
\[
\smash{\tau_{k_1k_2}(x,t;\theta_1,\theta_2)}=\exp\eta_2\cdot
 \smash{\bigl(1+\exp(-\eta_2)\cdot(1+\exp\eta_1)\bigr)}
\]
yields
\begin{multline}\label{SKdV4CaseLast}
u_{k_1k_2}(x,t;\theta_1,\theta_2)=A\,\cD_1\cD_2\Bigl[
 \bigl(k_2 x+\omega_2 t\bigr)
 +\log\bigl(1+\lambda_2\theta_1\theta_2\bigr)
 +\log\Bigl(1+\exp(-\eta_2)\cdot\bigl(1+\exp\eta_1\bigr)\Bigr)\Bigr]\\
{}\rightrightarrows -A\lambda_2=\text{const}.
\end{multline}
This value is the limit $\lim_{x\to+\infty}u_{k_2}(x,t)$ determined by the slow
soliton that now goes behind the fast. The constant does not alter under a
small finite motion left or right from the point $\teta_1=0$, where the peak
of the fast soliton would be located at any~$t\gg0$.

We have described the ``vanishing'' of the fast soliton. Now the visible slow
soliton is loaded with the virtual fast soliton.

It remains to note from~(\ref{SKdV4Case1st}--\ref{SKdV4CaseLast}) that
none of the solitons, neither the visible fast that becomes virtual, nor the
virtual slower soliton that becomes visible, acquires any phase shift.
The proof is complete.\hfill\qed

\medskip
Let us summarize our main result.

\begin{theor}\label{ThVirtualSoliton}
{\renewcommand{\theenumi}{\roman{enumi}}
\begin{enumerate}
\item
As $t\to-\infty$\textup{,} 
the time asymptote of the $n$-\/soliton~\eqref{nSoliton} 
for the super\/-\/KdV~\eqref{SKdV} is given by the \emph{fastest} soliton with
the maximal wave number~$k_{\text{\textup{max}}}$\textup{,} 
which is loaded with the
\emph{virtual} slower solitons that correspond to 
$k_2$\textup{,}\ $\ldots$\textup{,}\ $k_{\text{\textup{min}}}<
k_{\text{\textup{max}}}$. These virtual solitons are invisible at~$t\ll0$.
\item
As $t\to+\infty$\textup{,} the asymptote of the same solution 
$u_{k_1\cdots k_n}(x,t;\theta_1,\theta_2)$ for~\eqref{SKdV}
is the \emph{slowest} soliton with the minimal wave 
number~$k_{\text{\textup{min}}}$\textup{,} 
which is loaded with faster solitons 
determined by $k_{\text{\textup{max}}}$\textup{,}\ $\ldots$\textup{,} 
$k_{n-1}>k_{\text{\textup{min}}}$. 
These $n-1$ virtual solitons remain 
\textup{(}for~$k_2$\textup{,}\ $\ldots$\textup{,}\ $k_{n-1}$\textup{)} or
become \textup{(}respectively\textup{,} 
for~$k_{\text{\textup{max}}}$\textup{)} invisible at~$t\gg0$.
\item
None of the solitons\textup{,} neither the visible\textup{,} 
which goes behind the others\textup{,} nor
the virtual ones in front of it\textup{,} 
acquires any phase shift under the evolution
in time~$t\in(-\infty,+\infty)$.
\end{enumerate}%
}
\end{theor}

\begin{cor}
The iterated spatial and temporal limits of the pure imaginary first summand 
in~\eqref{nSoliton} are not permutable: as $x,t\to+\infty$, one has%
\footnote{Note that the limit in~$x$ in the right\/-\/hand side is taken
before the peak $x_n(t)$ of the slowest $n$-th soliton passes through
a point~$x$, which means that the exponentially vanishing tail of all solitons
has not arrived~yet.}
\[
\lim_{t\to+\infty}\lim_{x\to+\infty}\frac{\sum_{i=1}^n\lambda_i\alpha_i
  \exp\teta_i}{1+\sum_{i=1}^n\alpha_i\exp\teta_i}=
    \lambda\bigl(k_{\text{max}}\bigr) \neq
    \lambda\bigl(k_{\text{min}}\bigr) =
\lim_{\substack{x\to+\infty\\ x\gg x_n(t)}}
  \lim_{t\to+\infty}\frac{\sum_{i=1}^n\lambda_i\alpha_i
  \exp\teta_i}{1+\sum_{i=1}^n\alpha_i\exp\teta_i}.
\]
This also shows that the solution $u_{k_1\cdots k_n}(x,t;\theta_1,\theta_2)$
is not a Schwarz function in~$x$.   
Clearly, as $x\to-\infty$ at any~$t$, the limit of the first summand
in~\eqref{nSoliton} equals zero (we assume $k_i>0$ everywhere in the text).
The real coefficient of~$\theta_1\theta_2$ in~\eqref{nSoliton} tends to
$A(a)\,\Id/\Id x(\text{const})=0$ as $x\to\pm\infty$ at any~$t$.
\end{cor}

The initial limit as $x\to+\infty$ at any
finite time~$t$ is due to the visible fastest
soliton with the wave number~$k_{\text{max}}$, which is loaded with virtual
slower soliton(s). The spatial limit as $x\to+\infty$ for the out\/-\/going
state at large times $t\gg0$ is specified by the slowest soliton, which is
loaded with the virtual faster waves and which progressively occupies the
entire $x$-\/axis.

\begin{rem}
The $a=4$ super\/-\/KdV equation~\eqref{SKdV} possesses twice as many
Hamiltonian functionals and hence twice as many symmetries as the other two
cases $a\in\{-2,1\}$. Therefore there are twice as many commuting flows as in the hierarchies with $a=-2$ or $a=1$.
In particular, between the spatial translation $\dd/\dd x$
and the translation $\dd/\dd t$, 
which is determined by~\eqref{SKdV} in the $a=4$ super\/-\/KdV hierarchy, there is the
$N=2$ supersymmetric `Burgers' flow~\cite{Kiev2005},
\begin{equation}\label{N2Burgers}
\dot{u}=\cD_1\cD_2u_x+uu_x.
\end{equation}
Proposition~\ref{nSolState} and Theorem~\ref{ThVirtualSoliton} are literally
reproduced for
~\eqref{N2Burgers}.
The $n$-\/su\-per\-so\-li\-tons are~\eqref{nSoliton} with the following dispersion relations
in Hirota's $\tau$-\/function~\eqref{ZeroInteraction},
\begin{equation}\label{DispBurg}
A=2,\qquad \omega=-ik^2,\qquad \lambda=ik,\qquad \alpha\in\BBR. 
\end{equation}
\end{rem}

\paragraph*{Conclusion.}
The new $n$-\/soliton solutions~\eqref{nSoliton} satisfy $N=2$ super\/-\/KdV
equation~\eqref{SKdV} with $a\in\{1,4\}$ or, to be even
more precise, its bosonic
limit, which is the Krasil'shchik\/--\/Kersten system if~$a=1$.
Recall that the $t\to-\infty$ asymptote of our solutions~\eqref{nSoliton}
is given by the fastest soliton with the maximal wave
number~$k_{\text{max}}$, which is loaded with $n-1$ virtual slower solitons
ahead of it. We have proved that, looking at the initial profile and
recognizing it as a one\/-\/soliton solution up to any precision of
measurements, one can predict neither the out\/-\/going state nor even its
asymptote $-iAk_{\text{min}}$ at $t\gg0$ as $x\to+\infty$.

For an observer, the fast soliton ($k_1=k_{\text{max}}$)
is subject to a spontaneous decay.
The profile transforms into a collection of nonlinear\/--\/overlapping waves 
($k_1>k_2>\cdots>k_n$) that,
finally, constitute what one sees as another soliton with a different wave 
number $k_n=k_{\text{min}}$.
For a given number of solitons~$n\geq2$ and for \textit{a priori} given
wave numbers $k_{\text{min}}$ and $k_{\text{max}}$, this can be realized 
along any trajectory, which is a point in the configuration space
\[
\BBR_{+}^{n-2}\times\BBR^{n-1}\ni
 \bigl(k_2,k_2-k_3,\ldots,k_{n-2}-k_{n-1}\bigr)
  \times\bigl(\alpha_2,\ldots,\alpha_n\bigr),
\]
since over~$\BBC$ one can always rescale $\alpha_1=1$.

The absence of the phase shifts, supplemented with the presence of hierarchies
of conservation laws for Hamiltonian $N=2$ supersymmetric KdV 
equations~\eqref{SKdV}, demonstrates that the spatially extensive solitary
waves behave in elastic collisions as dimensionless particles 
  (as massive material points).
In this paper, we established the possibility of their spontaneous
decay, and we formulated the laws for transformation of observed particles
into the virtual ones and \textit{vice versa}.


{\small\paragraph*{Acknowledgements.}   
The authors thank P.\,Win\-ter\-nitz, A.\,S.\,So\-rin, M.\,A.\,Ne\-s\-te\-ren\-ko,
and J.\,van de Leur for stimulating discussions,  
and are grateful to the referee for remarks.
This research is supported by NWO (for A.\,K.) and 
NSERC (for V.\,H.);
partial financial support from the 5th International workshop
`Nonlinear Physics: Theory \& Experiment' 
is acknowledged.
A.\,K.\ thanks $\smash{\text{IH\'ES}}$ and
Department of Mathematics and Statistics, University of
Mont\-r\'eal for financial support and warm hospitality.

}

\end{document}